\documentclass[epj,twocolumn]{webofc}
\usepackage[varg]{txfonts}   
\usepackage{url,hyperref,breakurl}
\wocname{AtmoHEAD 2014}
\woctitle{EPJ Web of Conferences}

\begin{document}
\title{Strategy Implementation for the CTA Atmospheric Monitoring Program}

\author{
  Michele Doro\inst{1}\fnsep\thanks{\email{michele.doro@pd.infn.it}}
  \and
  Michael Daniel\inst{2}
  \and
  Raquel de los Reyes\inst{3}
  \and
  Markus Gaug\inst{4}
  \and
  Maria Concetta Maccarone\inst{5}
\\
\;for the CTA Consortium
}
  \institute{
    University and INFN Padova, via Marzolo 8, I-35131 Padova (Italy)
    \and
    Department of Physics, University of Liverpool, Liverpool, L69 7ZE (UK) 
    \and
    Max-Planck-Institut fu\"er Kernphysik, P.O. Box 103980, D-69029
    Heidelberg (Germany)
    \and
    Department of Physics, Universitat Aut\'onoma de Barcelona,
    Campus UAB, E-08193, Bellaterra (Spain)
    \and
    Istituto di Astrofisica Spaziale e Fisica Cosmica di Palermo,
    IASF-Pa/INAF, Via Ugo La Malfa 153, 90146 Palermo (Italy)
  }

\abstract{%
The Cherenkov Telescope Array (CTA) is the next generation
facility of Imaging Atmospheric Cherenkov Telescopes. It reaches
unprecedented sensitivity and energy resolution in very-high-energy
gamma-ray astronomy. CTA detects Cherenkov light emitted within an
atmospheric shower of particles initiated by cosmic-gamma rays or
cosmic rays entering the Earth's atmosphere. From the combination of images the Cherenkov light produces in the telescopes, one is able to infer the
primary particle energy and direction. A correct energy estimation can
be thus performed only if the local atmosphere is well characterized. The atmosphere not only affects the shower development itself, but also the Cherenkov photon transmission from the emission point in the particle shower, at about 10-20 km above the ground, to the detector. Cherenkov
light on the ground is peaked in the UV-blue region, and therefore molecular and aerosol extinction phenomena are important. The goal of CTA is to control
systematics in energy reconstruction to better than 10\%. For this reason,
a careful and continuous monitoring and characterization of the
atmosphere is required. 
In addition, CTA will be operated as an observatory,
with data made public along with appropriate analysis tools. High-level data quality can only be ensured if the atmospheric properties are consistently and continuously taken into account.

In this contribution, we concentrate on discussing the implementation
strategy for the various atmospheric monitoring instruments currently
under discussion in CTA. These includes Raman lidars and ceilometers, stellar
photometers and others available both from commercial providers and public research centers.
}
\maketitle

\section{Introduction}
\label{intro}
\vspace{-4mm}
\begin{figure}[h!t]
\centering
\includegraphics[width=0.95\linewidth]{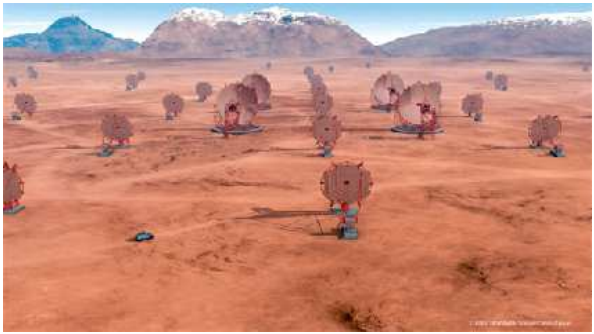}
\caption{\label{fig:cta} Artistic sketch of one of the two CTA sites. Several telescopes of different sizes will be deployed.}
\end{figure}
\vspace{-5mm}
Currently in its design stage, the Cherenkov Telescope Array (CTA, see
Fig.~\ref{fig:cta}) is an advanced facility for ground-based very-high-energy  gamma-ray
astronomy~\cite{bib:actis,acharya:2014}. It is an initiative to build
the next-generation ground-based very-high-energy gamma ray observatory covering the
energy range from a few tens of GeV up to over a hundred TeV with 
unprecedented sensitivity. The design of CTA is based on currently
available technologies and builds upon the success of the present
generation of ground-based Cherenkov telescope arrays (H.E.S.S., MAGIC
and VERITAS\,\footnote{\url{www.mpi-hd.mpg.de/hfm/HESS/,
  magic.mpp.mpg.de, veritas.sao.arizona.edu}}).  

Nowadays, the main contribution to the systematic uncertainties  
of imaging Cherenkov telescopes stems from the uncertainty in the height- and wavelength-dependent atmospheric optical properties for a
given run of data. Atmospheric quality affects the measured Cherenkov
yield in several ways: the air-shower development and Cherenkov light production, depending on the molecular density profile, the loss of 
photons due to scattering and absorption of Cherenkov light out of
the camera field-of-view, resulting in dimmer images, and the
scattering of photons into the camera, resulting in blurred images. 
Despite the fact that several supplementary instruments are currently
used to measure the atmospheric transparency together with Cherenkov
telescopes, their data are mainly used to retain good-quality
observation time slots, and only a minor effort 
has been made to routinely correct data with atmospheric
information~\cite{bib:nolan1,bib:dorner,bib:reyes,fruck:2013}.  
%~\cite{bib:nolan1,bib:dorner,bib:reyes,fruck:2013}.  
This situation will improve (MAGIC and H.E.S.S. are already
characterizing atmosphere to some extent, albeit not yet in a standard
way and not properly tested) with the CTA atmospheric monitoring and
calibration program~\cite{Doro:2013swa,Gaug:2014spie}. There are several goals behind this program. The
first is to increase the precision and accuracy in the energy and flux reconstruction through the
use of one or more atmospheric instruments. Secondly, a precise and
continuous monitoring of the atmosphere will allow for an increase of
the telescope duty cycle with the extension of
the observation time during hazy atmospheric conditions, which are
normally discarded in the current experiments because of the
uncertainty in the data reconstruction. Finally, a possible ``smart
scheduling'', i.e. an adaptation -- if required -- of the observation
strategy during the night that considers the actual atmospheric
condition can be activated with a precise monitoring program of the
atmospheric conditions.

This contribution is structured as follows. In Sec.~\ref{sec:influence}
we briefly discuss the influence of the atmosphere on IACT (Imaging
Atmospheric Cherenkov Telescope) data. In
Sec.~\ref{sec:instruments} we summarize the development activities for
the instrumentation and methods for the atmospheric monitoring program
in CTA. In Sec.~\ref{sec:strategy} we discuss the main tasks required
to the program as well as the implementation strategy. In
Sec.~\ref{sec:network} we discuss the possible usefulness of having
interchange of data between CTA and ground-based and satellite
atmospheric monitoring networks. We end with a summary.

\section{The Influence of the Atmosphere on IACT Data}
\label{sec:influence}
\vspace{-3mm}
\begin{figure}[h!t]
\centering
\includegraphics[width=0.98\linewidth]{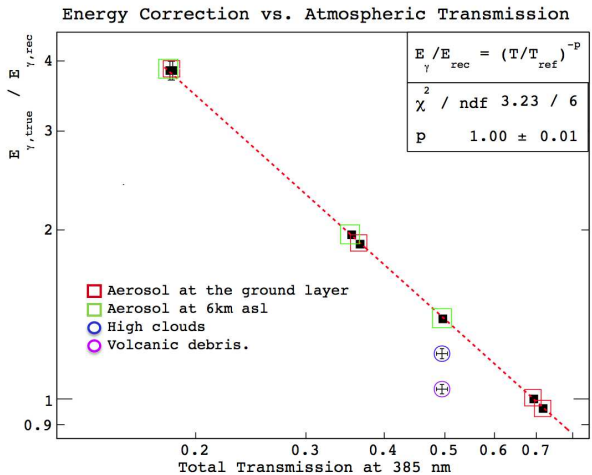} 
\caption{\label{fig:ecor}Energy correction factor as a function of the
  total atmospheric transmission at 385 nm. The red and green points show enhancements of the ground layer and low clouds, respectively, while the pink and dark green symbols refer to high clouds and volcanic debris. The linear dependence is valid only in
  the case of aerosol over-densities at low heights. $E_{\rm rec}$ is
  the reconstructed   energy from an initial Monte Carlo event with
  energy $E_\gamma$. $T_{\rm ref}$ is a scale transmission at 385~nm
  and   $-p$ the chi-square optimized index, $T$ is the transmission
  at   385~nm for each of the points in the plot. See text for
  additional details. Taken from~\cite{bib:garrido_phd}.}  
\end{figure}
\vspace{-3mm}
Although IACTs are normally placed at astronomical sites,
characterized by extremely good atmospheric conditions, the local
atmosphere is potentially influenced by phenomena occurring at tens to
thousands of kilometers away, and thus should be continuously
monitored. 
While the molecular content of the atmosphere varies very slowly at a
given location during the year, and slowly from place to place,
aerosol concentrations can vary on time-scales of minutes and travel
large, inter-continental, distances. Most of them are concentrated
within the first 3~km of the troposphere, with the
free troposphere above being orders of magnitude cleaner. 
Aerosol sizes reach from molecular dimensions to millimeters, and the particles
remain in the troposphere from 10 days to 3 weeks. The sizes are
strongly dependent on relative humidity. 
Different types of aerosol show characteristic size distributions, and
an astronomical site 
will always show a mixture of types, with one possibly dominant type at 
any given time and/or altitude.
Light scattering and absorption by aerosols needs to be described by
Mie theory or further developments of it, including  non-sphericity of
the scatterer. Aerosols generally have larger indexes of refraction
than that of water,  and some show also a small imaginary part. Contrary to the typical $\lambda^{-4}$ wavelength dependency of
Rayleigh-scattering molecules, aerosols show power-law indexes (the
so-called \textit{\AA ngstr\"om} coefficients) from  0 to 1.5, i.e. a
much weaker dependency on wavelength. 

In order to estimate the effect of different atmospheric conditions on
the image analysis of IACTs, we have simulated different molecular and
aerosol profiles for the MAGIC system, consisting of two
telescopes~\cite{bib:garrido_phd,garrido:2013}. Several aerosol
scenarios were simulated: $i)$ enhancements of the ground layer from a
quasi aerosol-free case up to  a thick layer which reduces optical 
transmission by 70\%, $ii)$ a cloud layer at the altitudes of 6~km, 
10~km (cirrus)  and 14~km~(volcano debris)~a.s.l., and $iii)$ a 6~km
cloud layer with varying aerosol densities. The main results can be
summarized in three points: 

\begin{enumerate}
\item Using adapted Monte Carlo simulation, energy and flux can be
  reconstructed right away, at
  only the expense of a larger energy threshold, which can be
  explained by the fact that the faint images of lower energy showers
  are absent with hazy atmospheres as fewer photons reach the
  ground. At energies above the new threshold, images observed under
  large impact distances, i.e. from coming from the shower halo, are 
 gradually lost, and a small degradation of the effective area is obtained;
\item In the case that the aerosol over-density or cloud is below the
  electromagnetic shower, a simple correction method can be used to
  restore correct energy and flux reconstruction with the simple use
  of standard Monte Carlo;
\item When the cloud or aerosol layer is at the shower development
  region or above, {\bf the total extinction is no longer a useful parameter};
\end{enumerate}

The last two points are shown in Fig.~\ref{fig:ecor} where the energy
correction is shown as a function of the total transmission at
385~nm. One can see that in the case that the aerosol over-density is located
close to the ground, i.e. below the electromagnetic shower
development, then the associated correction for the energy reconstruction is precisely obtained from the total transmission, while if the over-density is above, then
correlation is broken and {\bf height-resolving instruments are
  required}. This is the main motivation for the need of a Raman lidar
instrument for CTA. The main findings of this study must be also valid
for CTA, albeit the critical height of aerosol layers for the
simple linear energy correction to work out must lie considerably
lower, since showers of TeV energies penetrate further down the
atmosphere. 
Previous studies have been made
\cite{bib:bernloher,bib:nolan1,bib:dorner} for H.E.S.S. and
for the MAGIC mono system, however only for an increase of low-altitude
aerosol densities, and in \cite{bib:nolan2} for a reference
configuration of CTA, claiming a change in the spectral power-law
index of  gamma-ray fluxes, when atmospheric aerosol layers are
present. In our work, a very small dependence of the spectral index on
the aerosol densities could be obtained only with very dense
layers. See \cite{bib:garrido_phd,garrido:2013} 
for further details.
 
\section{Instruments and methods for the CTA Atmospheric Monitoring program}
\label{sec:instruments}
\subsection{Raman Lidars}
\vspace{-1mm}
%\begin{figure}[h!t]
%\centering
%\includegraphics[width=0.99\linewidth]{./lidars.eps}
%\caption{\label{fig:lidars} Lidars under development for CTA. (left)
%  Design by IFAE/UAB/LUPM (right) Design by CEILAP}
%\end{figure}
Several institutes in CTA are currently designing Raman lidar systems:
the Institut de F\`isica d'Altes Energies (IFAE) and the Universitat
Aut\`onoma de Barcelona (UAB), located in Barcelona (Spain), the LUPM
(Laboratoire Univers et Particules de Montpellier) in Montpellier
(France) and the CEILAP (Centro de Investigaciones Laser y sus
Aplicaciones) group in Villa Martelli (Argentina). The different
groups are designing independently prototypes with different
mechanical, optical and steering solutions. Generically, these lidars
are characterized  by large reflectors of $\geq$1.5 m diameter, powerful Nd:YAG
lasers operated at 2-3 wavelength (1054~nm and its first and second
harmonics), and an optical readout system based on at least 4 channels
(355 and 532 nm elastic and the corresponding two Raman lines). They are steerable and operated
mainly at night. More details of those devices can be found
in~\cite{Doro:2014csa}. 

In addition, two other
groups have shown interest in discussing new designs for Raman lidars,
one in Adelaide (Australia) that may provide a powerful wind-lidar,
and one from INFN (Italy) that will provide a smaller well-calibrated
Raman lidar for instrumental cross-check and site climatology
characterization~\cite{Valore:2014vba}.

\subsection{Ceilometer}
\vspace{-1mm}
A ceilometer is a device primarily used in the aviation industry for cloud
reporting and/or measuring the vertical visibility in harsh conditions. The
instrument makes use of the lidar technique, using low power infra-red pulsed 
diode lasers at high pulsing rates to provide eye-safe, reliable, fully
automatic 24/7 operation in all weather conditions. The continuous
operation of a ceilometer would not disturb the operation of the CTA. Cloud reporting range up to 
13 km (43,000 ft) and backscatter profiling over a full measurement range up to 
15 km (49,200 ft) is possible with a Vaisala CL51 model ceilometer. 
Advanced single-lens design provides excellent performance even at low
altitudes. In favourable conditions the backscattering profile can also be used
to monitor boundary layer structures~\cite{BL}.

\subsection{All-Sky Camera}
\vspace{-1mm}
%\begin{figure}[h!t]
%\centering
%%\includegraphics[width=0.99\linewidth]{./asc.eps}
%\caption{\label{fig:asc} All-sky camera housing (left) and scheme for
%  star recognition and optical transparency interpretation (right)}
%\end{figure}
The All-Sky Camera (ASC) is a passive non-invasive imaging system for night sky
atmosphere monitoring. The operation of the ASC would not affect the
measurement procedure of CTA. The ASC will determine clouds and an
overall atmospheric quality on the short term. In case of a partly cloudy night sky,
the cameras will identify the uncovered regions of the sky, which would
help to define those regions of the sky where sources can be observed without performance degradation~\cite{Mandat:2013}. 
%This way the higher
%productivity of the CTA observatory measurements could be improvedcan
%be achieved.  

\subsection{UVscope}
\vspace{-1mm}
%\begin{figure}[h!t]
%\centering
%%\includegraphics[width=0.99\linewidth]{./uvscope.eps}
%\caption{\label{fig:uvscope} Picture of the UVscope lunar photometer
%  developed by INAF and IASF}
%\end{figure}
The UVscope~\cite{Maccarone:2011}
 is a portable multi-pixel photon detector, NIST-calibrated in the lab, developed at INAF
IASF-Palermo (Italy) to support experimental activities in the high-energy
astrophysics and cosmic-ray field. The instrument, working in single
photon-counting mode, is designed to directly measure light flux in
the wavelength range 300-650 nm. Thanks to its features and
operational flexibility, the instrument can be used in a wide range of
applications. Currently, its primary application is in the framework of
the absolute calibration of the ASTRI SST-2M telescope
prototype~\cite{Maccarone:2014}. 
For CTA, the UVscope could be used to study the atmospheric transparency.

The atmospheric transparency can be evaluated by measuring the
absorption of the flux of a star, as a function of the air mass traversed,
which is related to the star altitude. The measurements of atmospheric
transparency are performed with the UVscope by following a star during its
path in the sky (tracking mode); in this way it is possible to
measure its luminosity as a function of the altitude, from which the sky 
transparency can be derived. The UVscope can also measure the diffuse
emission of the night sky background, NSB, in the same field of view
of a given telescope, as successfully performed at the Pierre Auger
Observatory during a several years campaign~\cite{Segreto:2011}. 
In the CTA framework, UVscope would follow the same pointing of a
given CTA telescope or group of telescopes.

\subsection{Photometric Robotic Atmospheric Monitor}
\vspace{-1mm}
%\begin{figure}[h!t]
%\centering`
%%\includegraphics[width=0.99\linewidth]{./fram.eps}
%\caption{\label{fig:fram} Picture of the FRAM Robotic Atmospheric Monitor}
%\end{figure}
The F/(Ph)otometric Robotic Atmospheric Monitor (FRAM) is a wide field astronomical imaging camera on a
robotic mount to monitor stars in the whole field of view of
the Cherenkov telescopes. From the measured light flux of the stars we
can determine the atmospheric extinction in this direction with high
spatial and temporal resolution without interfering with the Cherenkov
observations. The FRAM was successfully operated at the Pierre Auger
Observatory~\cite{BenZvi:2007uj}
and is currently being adapted to CTA. It consists of a
commercial robotic astronomical mount, a large-format CCD camera,
a photographic telephoto lens, a focuser for the lens, such as the
Rigel nStep focuser which 
mechanically couples to the manual focus wheel of the lens. 

While there are many probes capable of monitoring the state of the
atmosphere, stars have the advantage of being non-invasive, always
available and well understood. The FRAM samples
atmospheric properties with high precision at different directions
using well-characterized stellar fields and extrapolates these data
assuming spatial and temporal uniformity of the atmospheric
conditions. Secondly, it may perform detailed follow-up observations
covering wider, but still limited, areas of the sky immediately after
detection of particularly interesting air showers. The measurements of
the first kind are carried out using a narrow-field camera mounted
full 30-cm astronomical telescope (which is also used for other
observations, such as GRB follow-ups), while those of the second kind
use a wide-field camera connected to a photographic lens. 

\subsection{Molecular profile}
\vspace{-1mm}
The molecular profile is measured by radio sondes, but can be
approximated by global atmospheric data assimilation system, like the
GDAS~\cite{pao2012-2}. Its precision depends basically on the
proximity of standard radio sonde launch sites, like airports. As soon
as the sites for the CTA are selected, we will start a radio sonde
campaign of our own to cross-validate the local GDAS data and, upon
success, use these data for the (time-dependent) determination of the
molecular profile. 

\subsection{The Cherenkov Transparency Coefficient Method}
\vspace{-1mm}
Besides any other atmospheric monitoring devices placed on the
observatory site, the Cherenkov Transparency
Coefficient (CTC) will provide a way to estimate the atmospheric
transparency based on the information of the Cherenkov data only. This coefficient has
the advantage that is calculated using only observables and
calibration parameters (trigger rates, muon efficiency and mean camera
gain) from the Cherenkov data taken with the IACTs. 
The CTC is derived under the assumption that the zenith-corrected
single telescope trigger rate is dominated by cosmic-ray protons.
%characterized by a power law flux of spectral index $\Gamma$=-2.70. 
Therefore the trigger energy threshold of the telescope ($E_0$) is inversely
proportional to the average pixel gain $g$, the optical throughput of the telescope, parameterized by the muon efficiency $\mu$
and the atmospheric transparency, parameterized by a factor $\eta$ so
that $ E_0 \propto (\eta \cdot \mu \cdot g )^{-1} $. 
Random fluctuations in the trigger of a single telescope are removed
by selecting only the read-out events for which at least 2 telescopes
are triggered in coincidence. This read-out trigger will therefore
depend on the number of active telescopes so the averaged trigger over
all $N$ active telescopes is calculated and rescaled by a factor of
$k_N$ that depends on the telescope multiplicity.  

Different correlation studies have been carried out between
the CTC and independent measurements from diverse atmospheric monitoring
instruments~\cite{Hahn:2014}.
A large correlation factor of $\sim$ 0.85 at blue wavelengths
($\lambda=443$~nm) has been found between the aerosol optical
depth measurements performed by the MISR satellite and the CTC.
The same order of magnitude has been found in the correlation with lidar and
Radiometer systems placed on the H.E.S.S. 
site. It does not
(yet) provide information about the height profile of the atmospheric
disturbances, which would be required for a full data correction,
though.  
%(23$^\circ$16'18'' S, 16$^\circ$30'00'' E,
%1800\,m a.s.l.). 
These positive correlation results prove the 
capability of this coefficient to characterize the aerosol optical depth
affecting the Cherenkov observations.
Since the CTC is extracted from the Cherenkov data 
used in the IACT analysis, it is easy to study the relation between 
the atmosphere quality and the systematic errors introduced in the 
spectral analysis results due to a bad calibration of the
atmosphere~\cite{Hahn:2014}.
Therefore, the CTC is currently used in H.E.S.S. as a data quality selection
criteria, establishing a criteria for the maximum allowed systematic error in the 
spectral analysis results of $\sim20$\%.
The relation of the CTC not only with the Cherenkov analysis
systematics but also with independent atmospheric instrumentation
on-site opens the possibility of an automatic interaction 
with a central autoscheduler concerning the atmospheric
conditions for observations. 
\section{Strategy Implementation for CTA Atmospheric Monitoring program}
\label{sec:strategy}
\vspace{-1mm}
As discussed above, an ensemble of instruments is foreseen to be
installed at the site(s) of CTA for the Atmospheric
Monitoring. Obviously, these instruments, when properly characterized
and inter-connected, have various uses. In this section, we outline the
main tasks we expect from those instruments, as well the strategy
implementation. 

\subsection{TASK-1. Climatology Build-up} \vspace{-1mm}This task deals with studies
carried out as soon as the site decision is made, even before the
first telescopes are installed. Its outcome will be needed for the
decision as to whether data models like GDAS or SENES are reliable enough,
and which part can be used to replace direct measurements. During this
time, also a profound knowledge of the aerosol mass distributions,
their occurrence and variability should be obtained. In addition, we use
this phase to gain full knowledge about the correction of the
telescope data for atmospheric influences. During this time, all
available atmospheric monitoring instruments should be in place, and
their interplay with the telescopes be studied, as well as their
contribution to the correction factors of the telescope data.  

Among the instruments foreseen for this task,
we propose as primary (P) and secondary (S) -- in terms of relevance
--:
\begin{itemize}
\item (P) Commercial weather stations, satellites and local historical
  data, national radars, radio sondes, INFN Raman lidar, ASC 
\item (S) Ceilometers, All-sky Camera, UVscope and FRAM, Raman lidars
\end{itemize}

Commercial weather stations are meant to generally characterize the
current situation of the site in terms of humidity, temperature,
wind speed and direction and pressure and correlations of rain fall and thunderstorm activity with the previous parameters. However, the collection of
multi-annual data and their correlation with available satellite
data, historical data and with data from possible closeby national
radar can reinforce the knowledge of long-term behavior at the site,
and therefore allow to better model any current situation and allow
to make predictions. These instruments and methods should therefore be
installed/implemented at the site as soon as possible. In
addition, the use of a sun photometer inside global networks like
AERONET, will allow not only to validate and disseminate our data, but
to have a more precise global scenario. Finally, radiosondes are
extremely important in the first years to not only monitor the
situation in-situ, but also collect valuable information of
the atmospheric vertical structure. The launches of balloons are only
foreseen for the first years of CTA, probably before the start of
operation. The INFN Raman lidar comes from the ARCADE
project~\cite{bib:ICRC_Arcade} and is a well-calibrated lidar. It
has a 20~cm diameter mirror and is more easily portable than other Raman
lidars under development for CTA. It will be brought to the site as
soon as it becomes available with the additional task of cross-calibrating the
final Raman lidars for the site. The ASC is in a mature stage and will
be installed at the site soon.

However, additional instruments can complement the acquisition of a
climatology of the site, as well as themselves being better
calibrated and commissioned. The ceilometer will allow to retrieve
information on the cloud presence 
and distribution in the sky both day and night. Stellar photometers
data like that from the FRAM will allow to commission
those devices as well as know the precise extinction in given
directions. Finally, Raman lidar data, compared with ASC, FRAM and
UVscope data will allow to improve the overall quality and precision
of the results.

\subsection{TASK-2. Off-line data selection} \vspace{-1mm}The goal of the CTA
atmospheric calibration is to minimize any loss of data due to
non-optimal atmospheric conditions during data selection. This
shall be achieved through smart scheduling according to the needs of
the physics analysis behind the observation, in combination with the
available atmospheric conditions in a given pointing direction. If the
conditions do not allow such a strategy, then offline data correction
or even rejection will gain an important role. 

Among the instruments and methods that can be used for such a task are:
\begin{itemize}
\item (P) Raman lidars, UVscope and FRAM and CTC
\item (S) All-sky camera
\end{itemize}
Raman lidar data will provide spatially resolved information, but the
information coming from integrated optical depth from the FRAM and the
UVscope will allow to improve the precision. Finally, from the data
of the telescope, we can cross-check the above information with the
CTC. 

Even if the ASC is an all-sky instrument, it also serves the purpose
of measuring the integral extinction from a certain direction in the
sky and therefore can complement the above information.

\subsection{TASK-3. Off-line data correction} \vspace{-1mm}Data taken during
non-optimal atmospheric conditions can be corrected by the use of adapted Monte
Carlo simulations or clever algorithms that make use of atmospheric
information~\cite{fruck:2013}. As a result, not  only the energy and flux can be correctly
retrieved for several situations, but also the overall duty cycle of
the experiment is increased. This task is similar to Task-2, but
requires more precision, as well as more complicated analysis
algorithm, interactions with the data reconstruction pipeline, the
Monte Carlo simulations, etc.

The instruments and methods for this task are therefore the same as for Task-2,
\begin{itemize}
\item (P) Raman lidars, UVscope and FRAM and CTC
\item (S) All-sky camera
\end{itemize}
The reason we discuss this task separately is that its complexity is higher and requires the densest data coverage achievable, as well as robustness of the result, in order
to reduce and control the final systematics quoted in the scientific
publications. 

\subsection{TASK-4. On-line smart scheduling} This section deals with
the instrumentation and procedure to assure that the CTA points always
to sources under conditions which permit to carry out the relevant
physics analysis afterwards. Depending on the altitude and thickness of
aerosol/cloud layers, the energy threshold as well as angular and
energy resolution may be degraded, until an observation under the
given conditions would not make sense any more. Given that CTA will
deploy instrumentation that  allows to predict these parameters, a 
smart scheduling program will decide, at any time, to select only
observations fulfilling the criteria on threshold and resolution. For
this, we therefore work with all-sky instruments, rather than
instruments that classify the   atmosphere in the pointing direction
as in Task-2 and -3. Of importance are:  
\begin{itemize}
\item (P) All-sky Camera, Ceilometer
\item (S) Raman lidars, FRAM, UVscope and CTC
\end{itemize}
It is true however, that while the above instruments provide
information on the whole sky, the precise knowledge on the current
observation is provided by the pointing instruments. For example,
Raman lidars can provide a precise estimation on the current energy
threshold of the experiment, and in case of peculiar objects this
could suggest to move to another source in the sky. 

\subsection{TASK-5. Weather now-cast, fore-cast, alerts and
  protection} 
\vspace{-1mm}Atmospheric instruments provide online information on the weather and
can be used to make short-term forecasts that are useful not only for 
data-taking, but for planning trips, interventions, etc.
Atmospheric instruments can be used to forecast or early  detect
possible risky situations  for the experiment, like rain, lightning,
thunderstorms, etc, with the possibility to raise alerts and take
decisions over subsystems. Of primary importance are:
\begin{itemize}
\item (P) Commercial weather stations, national weather radars, satellites, remote rain sensors, lightning sensors, ceilometers
\end{itemize}

\section{Networks of atmospheric instruments}
\label{sec:network}
Ground-based atmospheric monitoring networks provide input
to global atmospheric models. Networks are important also because they
impose standardized algorithms and reduce bias. For a more detailed
discussion, we refer the reader to~\cite{Louedec:2014soa}. 

The biggest
atmospheric network for chemical composition
of the atmosphere is the Global Atmospheric Watch (GAW)
programme of the World Meteorological Organisation
(WMO)~\cite{GAW}. 
%This network is composed of more than 400 
%surface-based stations, including 29 elements called ''global
%stations'' where all measurements required
%in the GAW programme are operated~\cite{Louedec:2014soa}. 
Regarding aerosol radiative properties, there are other
global networks, like the AErosol RObotic NETwork
  (AERONET)~\cite{AERONET_website}. 
%This 
%network, managed by the NASA and the CNRS, monitors the total aerosol
%optical depth among other parameters. 
Other smaller networks exist
like GAW Aerosol LiDAR
Observations Network (GALION) and sub-networks as
the European Aerosol Research LiDAR NETwork
(EARLINET) or the Micro-Pulse
LiDAR NETwork (MPLNET)~\cite{MPLNET_website}. Regarding clouds,
CLOUDNET monitors the cloud coverage and its vertical
structure~\cite{CLOUDNET_website}.   

Satellites are also available to offer
accurate measurements in regions not covered (yet) by ground-based
weather stations. Satellites can be 
geostationary or polar. Satellites typically measure the cloud
emissivity in infrared or sunlight 
scattered by aerosols or clouds, or record backscattered
light. In this category, we mention  CALIPSO, MODIS
GOES~\cite{CALIPSO}, among others. Data from satellites are usually available
publicly one year later or once the mission has ended and therefore can
be used easily to evaluate atmospheric conditions of a site candidate
for an astroparticle physics experiment. 

It is currently under investigation what kind of interaction the CTA
instrumentation can  sustain with these networks. From one side, CTA
can obviously profit from the wealth of data provided by those
networks. On the other side, these networks could be interested in
obtaining data from the Atmospheric Monitoring instruments of CTA,
especially because these will be located at remote sites, normally not
covered by ground-based installations.

\section{Conclusion}
\label{sec:conclusion}
CTA will constitute the leading project in high energy gamma-ray
astronomy in the future decades. It can provide excellent data quality
provided the atmosphere is well determined during data-taking. To
achieve this goal, an ensemble of instruments is currently planned for the CTA sites that can perform pointed and all-sky
observation with several different tasks including providing a site
climatology, data selection and correction, smart scheduling as well
as climate forecast and instrument protection. 

\footnotesize{%
\subsubsection*{Acknowledgments}
\vspace{-2mm}
We gratefully acknowledge support from the agencies and organizations
listed under Funding Agencies at \url{http://www.cta-observatory.org/} 
}

\end{document}